\documentclass[prl, twocolumn, showpacs, english, preprintnumbers, amsmath, amssymb, superscriptaddress, aps]{revtex4-1}
\usepackage{latexsym}
\usepackage{epsfig,psfrag}
\usepackage{dcolumn}
\usepackage{bm}
\usepackage{graphicx}
\usepackage{color}
\usepackage{esint}
\usepackage{babel}
\usepackage{amsfonts}
\usepackage{enumerate}
\begin{document}
\title{Electric dipole induced universality for Dirac fermions in graphene} 

\author{Alessandro De Martino}
\affiliation{Department of Mathematics, City University London, 
London EC1V 0HB, United Kingdom}

\author{Denis Kl\"opfer}
\affiliation{Institut f\"ur Theoretische Physik, 
Heinrich-Heine-Universit\"at, D-40225 D\"usseldorf, Germany}

\author{Davron Matrasulov}
\affiliation{Turin Polytechnic University in Tashkent, 17 Niyazov Str., 
100095 Tashkent, Uzbekistan}

\author{Reinhold Egger}
\affiliation{Institut f\"ur Theoretische Physik, 
Heinrich-Heine-Universit\"at, D-40225 D\"usseldorf, Germany}

\date{\today}

\begin{abstract}
We study electric dipole effects for massive Dirac fermions in graphene and related
materials.  The dipole potential accomodates towers of infinitely
many bound states exhibiting a universal Efimov-like scaling hierarchy. 
The dipole moment determines the number of towers, but there is always
at least one tower.  The corresponding eigenstates show a
characteristic angular asymmetry, observable in tunnel spectroscopy.  
However, charge transport properties inferred
from scattering states are highly isotropic.
\end{abstract}

\pacs{72.80.Vp, 73.22.Pr, 71.15.Rf} 

\maketitle

\textit{Introduction.---}Close to the neutrality point, the 
quasiparticle excitations in a graphene monolayer are two-dimensional (2D) 
Dirac fermions \cite{rmp1}, where a gap $\Delta$ 
can be opened, e.g., by strain engineering \cite{strain}, 
spin-orbit coupling \cite{soi}, strong electron-electron 
interactions \cite{rmp2}, substrate-induced superlattices 
\cite{pono,levitov2}, or in a ribbon geometry \cite{rmp1}.
Graphene thus provides experimental access to relativistic quantum 
effects such as supercriticality, where a Coulomb impurity of charge 
$Q=Ze$ accomodates bound states that 'dive' into the filled Dirac sea 
for $Z>Z_c$ \cite{rmp2,novikov,pereira,levitov,gamayun,denis,exp1,exp2,exp3}. 
While $Z_c\approx 170$ is normally prohibitively large \cite{greiner,popov}, 
the smaller value $Z_c\approx 1$ in graphene has revealed
supercriticality in tunneling spectroscopy \cite{exp2,exp3}, where
the impurity was created by pushing together 
charged Co \cite{exp1} or Ca \cite{exp3} adatoms with a 
STM tip.  The charge $Q$ of the resulting 
cluster can be tuned by a local gate voltage.  Arranging suitably 
charged clusters ('nuclei') on graphene, one may then design
'molecules' in an ultrarelativistic regime otherwise unreachable. 

\begin{figure}
\centering
\includegraphics[width=8.5cm]{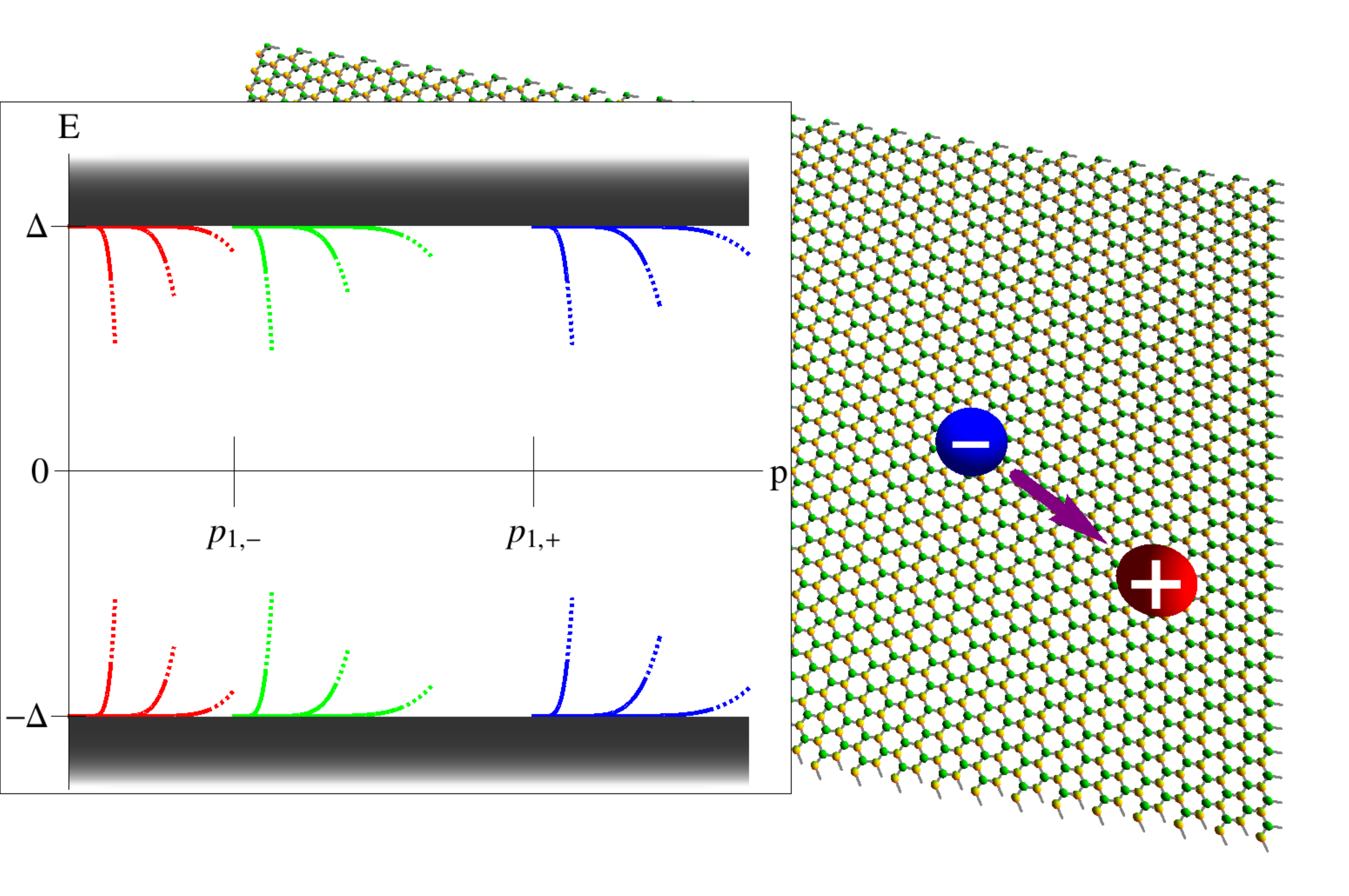}
\caption{\label{fig1} Sketch of the spectrum vs dipole moment $p$
for Dirac fermions with gap $\Delta$ in a dipole potential.  For $|E|>\Delta$,
we have scattering states [Eq.~(\ref{scatt})].  Bound states inside the gap are arranged in 
$(j,\kappa=\pm)$ towers.  Such a tower exists only when $p>p_{j,\kappa}$,
with $p_{0,+}=0$ and $p_{j>0,\kappa}$ in Table~\ref{table11}.
Bound states satisfy the Efimov scaling law [Eq.~(\ref{efimovscaling})],
where both gap edges are accumulation points. 
The background shows the schematic setup.}
\end{figure}

Here we predict universal quantum effects, different from 
supercriticality, for Dirac fermions in the $1/r^2$ \textit{dipole} 
potential of two oppositely charged ($\pm Q$) nuclei at distance $d$,
with electric dipole moment $p=Qd$.  Surprisingly, the Dirac dipole problem 
has not been discussed so far, presumably because of the 
lack of heavy anti-nuclei preventing its realization in atomic physics.  
However, it could be directly studied using STM spectroscopy in graphene 
\cite{exp1,exp2,exp3}. A similar $1/r^2$ potential also describes
 conical singularities \cite{wrinkle}.  Our main results are as follows, 
cf.~Fig.~\ref{fig1}.
(i)  The spectrum is particle-hole symmetric.  Bound states inside the gap, 
$E=\pm(\Delta-\varepsilon)$ with binding energy $\varepsilon \ll \Delta$,
come in ($j,\kappa$) towers of definite 'angular' quantum 
number, $j=0,1,2,\ldots$, and parity $\kappa=\pm$ (with $j+\kappa\ge 0$). 
The $(j,\kappa)$ tower is only present if 
the dipole moment exceeds a critical value, $p>p_{j,\kappa}$, 
but then contains infinitely many bound states.  
Since $p_{0,+}=0$, there is at least one such tower.  The lowest-lying 
finite  $p_{j,\kappa}$ are listed in Table \ref{table11}, 
with excellent agreement between two different derivations.
(ii) Bound states in the same tower obey the scaling hierarchy
\begin{equation}\label{efimovscaling}
\frac{\varepsilon_{n+1}}{\varepsilon_{n}}= e^{-2\pi/s_{j,\kappa}}, \quad n=1,2,\ldots
\end{equation} 
where for $p$ close to (but above) $p_{j,\kappa}$, 
\begin{equation}\label{ssdef}
s_{j,\kappa}(p) \simeq 
\left\{ \begin{array}{ll} \sqrt{2} p\Delta, & (j,\kappa)=(0,+),\\
 \alpha \sqrt{(p-p_{j,\kappa})\Delta},& j>0, \end{array}\right.
\end{equation}
with $\alpha\approx 0.956$.  As $n\to \infty$, all bound states
approach 
one of the gap edges as accumulation point.  
Equation (\ref{efimovscaling}) agrees with
the universal Efimov law for the binding energies of three 
identical bosons with short-ranged particle interactions \cite{efimov,efimov2,gogolin}.  
(iii) Numerical diagonalization of the Dirac equation in a finite disc geometry 
indicates that as $p$ increases, the bound states approach $E=0$ without ever reaching it.
The absence of zero modes is also shown analytically. 
(iv) The scattering state for $|E|\gg \Delta$ implies an isotropic
transport cross-section, such that charge transport is independent
of the angle between current flow and dipole direction.

\begin{table}
\caption{
\label{table11}
The lowest few finite critical dipole moments,  where
 $p_{j,\kappa}^{\rm M}$  follows from the Mathieu eigenvalues, 
Eq.~(\ref{pjdef}), and $p_{j,\kappa}^{\rm AK}$ from solving
the two-center problem, Eq.~(\ref{akdef}).}
\begin{tabular}{|l||l|l|}
\hline 
$(j,\kappa)$ & $p^{\rm M}_{j,\kappa} \Delta$ & 
$p_{j,\kappa}^{\rm AK}\Delta$ \\ \hline
$(1,-)$ &  1.89492  & 1.88805 \\
$(1,+)$ &  5.32466  & 5.32565 \\
$(2,-)$ &  10.4819  & 10.4820 \\
$(2,+)$ &  17.3571  & 17.3572 \\
$(3,-)$ &  25.9511  & 25.9512 \\
$(3,+)$ &  36.2639  & 36.2640 \\ \hline
\end{tabular}
\end{table}

\textit{Model.---}We study 2D Dirac fermions with a mass gap $\Delta$.
With $\hbar=e=1$ and Fermi velocity $v=1$, the Hamiltonian is 
\begin{equation}\label{eq1}
H= (-i\partial_x) \sigma_x+(-i\partial_y)\sigma_y +\Delta \sigma_z + V.
\end{equation}
In graphene, the two components of the spinor, $\Psi=(\phi,\chi)^T$,
correspond to the two sublattices, where the Pauli matrices 
$\sigma_{x,y,z}$ act in this space and we consider a single $K$ 
point and fixed spin projection \cite{rmp1}. 
Equation (\ref{eq1}) also describes 'molecular graphene' with
CO molecules deposited on a copper surface \cite{gomes},
and the surface states of topological insulators 
like Bi$_2$Se$_3$ or Bi$_2$Te$_3$ \cite{hasan}.  
Assuming two oppositely charged nuclei at $x=\pm d/2$, the  potential reads
\begin{equation}\label{twocenter}
V(x,y)= \frac{p/d}{\sqrt{(x+d/2)^2+y^2}}-
\frac{p/d}{\sqrt{(x-d/2)^2+y^2}},
\end{equation}
where $p=Qd$ is the dipole moment \cite{foot1}.
For equal charges 
in the two-center potential (\ref{twocenter}), 
similar physics as for a single impurity is found \cite{gusynin}.
In polar coordinates the Dirac equation reads
\begin{equation}\label{diracdip}
\left( \begin{array}{cc} V +\Delta-E &
 e^{-i\theta} (-i\partial_r - \frac{1}{r} \partial_\theta ) \\ 
 e^{i\theta} (-i\partial_r + \frac{1}{r} \partial_\theta ) & 
V - \Delta-E\end{array} \right) \left( \begin{array}{c}
\phi \\ \chi \end{array} \right)=0.
\end{equation}
Far away from the nuclei, $r\gg d$, Eq.~(\ref{twocenter}) 
is well approximated by the point-like dipole form
\begin{equation}\label{pointdipole}
V_d(r,\theta)= - \frac{p \cos\theta}{r^2}.
\end{equation}
The $r\to 0$ singularity implies that Eq.~(\ref{diracdip}) for $V=V_d$
requires regularization to avoid the usual fall-to-the-center 
problem \cite{popov2}. To that end one may resort to $V$ in Eq.~(\ref{twocenter}), 
but simpler regularization schemes are also possible, see below.
For nonrelativistic Schr\"odinger fermions, 
the dipole captures bound states only above a 
finite critical dipole moment in three dimensions (3D)
 \cite{AK,davron,dipol0,dipol1,num-efimov}.  However,
a dipole binds states for arbitrarily small $p$ in the 2D
Schr\"odinger case \cite{dipol1}.  

\textit{Particle-hole transformation.---}The 
Hamiltonian (\ref{eq1}) with $V$ in Eq.~(\ref{twocenter}) is mapped to 
$UHU^\dagger= -H$ by the 
unitary transformation 
$U=\sigma_x {\cal R}_x$, 
with ${\cal R}_x$ the reflection $x\to -x$. 
An eigenstate $\Psi_E(x,y)$ at energy $E$ is mapped to another 
eigenstate at energy $-E$,
\begin{equation}\label{unitary}
\Psi_{-E}(x,y) = U \Psi_E(x,y)= \sigma_x \Psi_E(-x,y).
\end{equation}
Hence all solutions to Eq~(\ref{diracdip}) 
come in $\pm E$ pairs. It is then sufficient to study
$E>0$ only, with the $-E$ partner state following from
Eq.~(\ref{unitary}).  The dipole moment  sign is
also irrelevant, and $p>0$ below.  

\textit{Near the band edges.---}We first consider 
Eq.~(\ref{diracdip}) for energies close to the band edge,
 $E=-\Delta+\varepsilon$  with $|\varepsilon|\ll \Delta$,
where $\varepsilon>0$ corresponds to 
bound states inside the gap and $\varepsilon<0$ to continuum states.
For $p \ll d^2 \Delta$, the upper spinor component stays always 'small', 
$\phi \simeq \frac{1}{2\Delta} e^{-i\theta}
\left(i\partial_r + \frac{1}{r}\partial_\theta\right)\chi$,
and Eq.~(\ref{diracdip}) leads to an effective 
Schr\"odinger equation  for the lower spinor component,
\begin{equation}\label{schr}
\left( -\frac{1}{2\Delta} \nabla^2- V+\varepsilon \right) \chi = 0, 
\end{equation}
with the 2D Laplacian $\nabla^2$.  We proceed with the 
potential $V=V_d$ in Eq.~(\ref{pointdipole}), where Eq.~(\ref{schr}) 
is solved by the ansatz $\chi(r,\theta) = R(r) Y(\theta)$.  
With separation constant $\gamma$, the angular function 
satisfies an $\varepsilon$-independent Mathieu equation,
\begin{equation}\label{angular}
\left(\frac{d^2}{d\theta^2} + \gamma-2p\Delta\cos\theta \right)Y(\theta)=0,
\end{equation}
which admits $2\pi$-periodic solutions only for characteristic 
values $\gamma=\gamma_{j,\kappa}(p)$, where 
$\kappa=\pm$ is the parity, i.e., 
$Y_{j,\kappa}(-\theta)= \kappa Y_{j,\kappa}(\theta)$, and
due to the anisotropy, $j=0,1,2,\ldots$ differs from 
conventional angular momentum, with $j+\kappa\ge 0$.
Using standard notation \cite{gradst,abramowitz}, 
the solutions to Eq.~(\ref{angular}) are expressed in terms of 
Mathieu functions ${\rm ce}_{2j}$ and ${\rm se}_{2j}$, 
with eigenvalues $a_{2j}$ and $b_{2j}$, respectively, 
\begin{eqnarray}\label{mathieusol} 
Y_{j,+}(\theta) &=& {\rm ce}_{2j}\left(\frac{\theta}{2}, 4p\Delta
\right) ,\quad \gamma_{j,+}= \frac{1}{4}
a_{2j}(4p\Delta), \\  \nonumber
Y_{j,-}(\theta) &=& {\rm se}_{2j}\left(\frac{\theta}{2}, 4p\Delta
\right),\quad \gamma_{j,-}=\frac{1}{4} b_{2j}(4p\Delta).
\end{eqnarray}
The characteristic values are ordered as 
$\gamma_{0,+}<\gamma_{1,-}< \gamma_{1,+}
<\gamma_{2,-}<\ldots$ for given $p$.
With $\gamma=\gamma_{j,\kappa}(p)$, the radial equation reads
\begin{equation}\label{radial}
\left(\frac{d^2}{dr^2} + \frac{1}{r} \frac{d}{dr} - \frac{\gamma}{r^2}-
2\Delta \varepsilon \right) R(r) = 0.
\end{equation}
To regularize the fall-to-the-center singularity, 
we impose the Dirichlet condition
$R(r_0)=0$ at a short-distance scale $r_0\approx d$ \cite{foot2}.
We show below that this regularization does not affect 
universal spectral properties such as the Efimov law (\ref{efimovscaling}).

\textit{Efimov scaling.---}Let us now look for bound states, $\varepsilon>0$.  
The solution of Eq.~(\ref{radial}) decaying for $r\to \infty$
is the Macdonald function $K_{\sqrt{\gamma}} 
\left(\sqrt{2\Delta \varepsilon}\ r\right)$ \cite{gradst}, and
$R(r_0)=0$ then yields an energy quantization condition
within each ($j,\kappa$) tower. Thereby the binding energies,
$\varepsilon_{n,j,\kappa}=z_{n}^2 /(2\Delta r_0^2)$, 
are expressed in terms of the positive zeroes, 
$z_1>z_2>\ldots>0$, of $K_{\sqrt{\gamma_{j,\kappa}}}(z)$.   
Since only $K_{is}(z)$ (with imaginary order)
has zeroes \cite{gradst}, bound states require $\gamma_{j,\kappa}(p)<0$.
This condition is satisfied for $p>p_{j,\kappa}$ with
\begin{equation}\label{pjdef}
\gamma_{j,\kappa}\left (p_{j,\kappa}\right)=0.
\end{equation}
The lowest few $p_{j>0,\kappa}$ resulting from Eq.~(\ref{pjdef}) are 
listed in Table \ref{table11}. With increasing dipole moment, each time 
that $p$ hits a critical value $p_{j,\kappa}$, a new infinite tower of bound 
states emerges from the continuum.  Since $\gamma_{0,+}(p)<0$ for all 
$p$ \cite{abramowitz}, we find $p_{0,+}=0$: at least
 one tower is always present.  
Explicit binding energies follow from the small-$z$ 
expansion of $K_{is}(z)$ \cite{gradst}. With the positive 
numbers $s_{j,\kappa}(p)= \sqrt{-\gamma_{j,\kappa}(p)}$ 
for $p>p_{j,\kappa}$,  see Eq.~(\ref{ssdef}),
we obtain 
\begin{equation}\label{spectrum}
\varepsilon_{n,j,\kappa} = \frac{2}{\Delta r_0^2}
 e^{\varphi(s_{j,\kappa})}
 e^{-2\pi n/s_{j,\kappa}}, 
\end{equation}
where $\varphi(s) = (2/s)\ {\rm arg} \Gamma(1+is)$.
This becomes more and more accurate as $n$ increases.  For $n\to \infty$, 
using particle-hole symmetry, the energies 
accumulate near both edges, $\varepsilon_{n}\to 0$.   Importantly, 
Eq.~(\ref{spectrum}) implies the Efimov scaling law announced in 
Eq.~(\ref{efimovscaling}).  This relation has its origin in the
\textit{large-distance} behavior of the dipole potential,
and is thus expected to be independent of short-distance 
regularization issues.  A similar behavior has been predicted for the 
quasi-stationary resonances of a supercritical
Coulomb impurity in graphene \cite{levitov,gamayun}, and for 3D Schr\"odinger 
fermions \cite{AK,num-efimov}. 

\begin{figure}
\centering
\includegraphics[width=8.7cm]{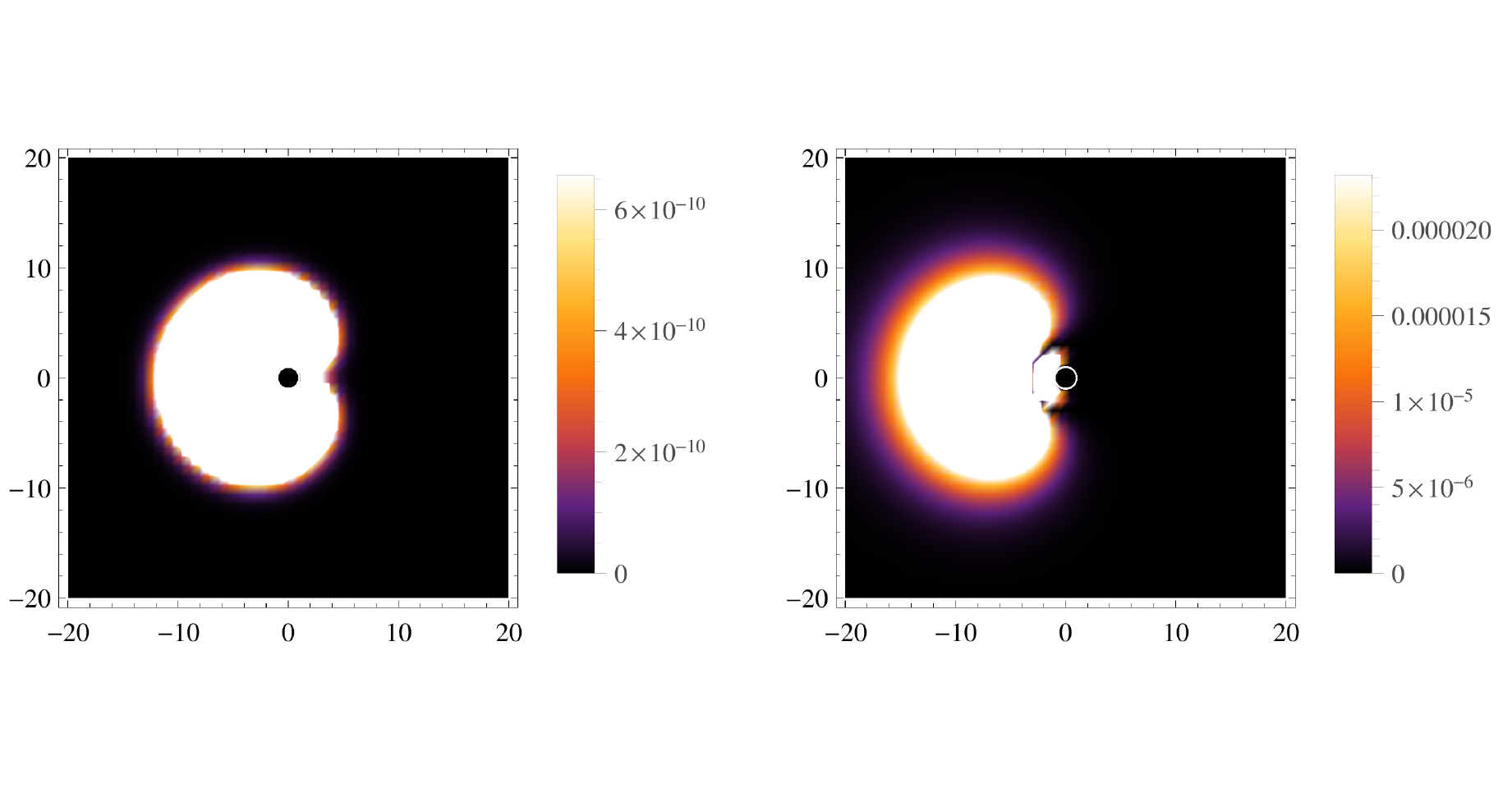}
\includegraphics[width=9cm]{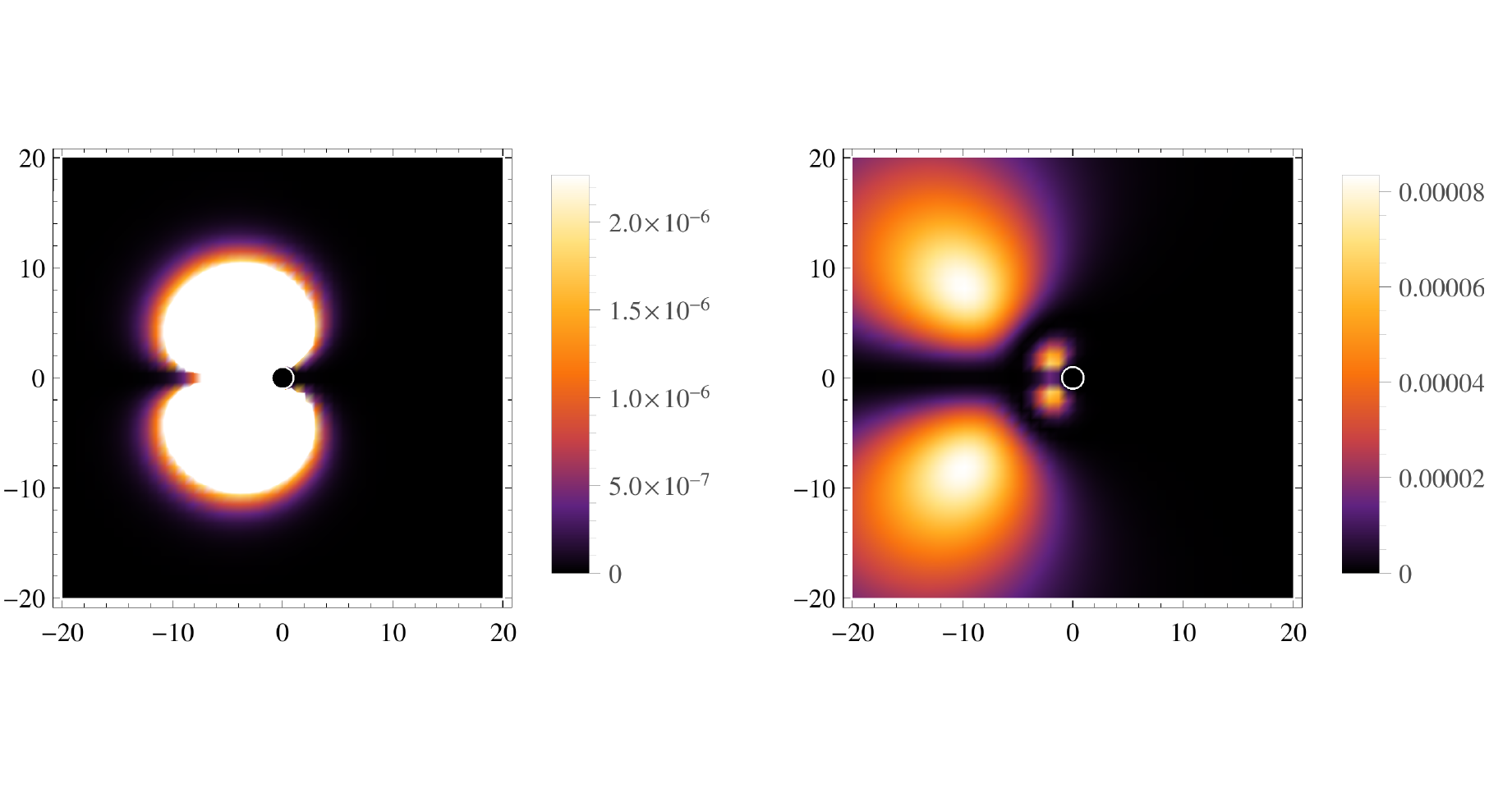}
\caption{\label{fig2} Color-scale plot of $|\Psi(x,y)|^2$, in the $x$-$y$ plane,
where $x$ and $y$ are in units of $r_0=d/2$, for several 
bound states with $p\Delta=5$. For $r<r_0$, the density vanishes
due to the Dirichlet condition.
The upper part shows the $n=1$ (left) and the $n=2$ (right) radial states
in the $(0,+)$ tower.  The lower part shows the same but
 for the $(1,-)$ tower.  }
\end{figure}

\textit{Tunneling density of states.---}The above solution also 
yields the probability density $|\Psi(r,\theta)|^2$, 
which is probed by the local tunneling density of states 
when the energy matches the respective bound
state energy, and can be measured in STM spectroscopy experiments
\cite{exp1,exp2,exp3}.  Figure \ref{fig2} shows typical results
for the two lowest hole-like radial bound states ($n=1,2$)
in the $(0,+)$ and $(1,-)$ tower, respectively.  The pronounced asymmetry 
along the $x$-direction is due to the Mathieu functions 
in Eq.~(\ref{mathieusol}) and is a characteristic 
feature to look for in experiments. 
The reflected ($x\to -x$) profile is found for the 
electron-like partner at energy $+|E|$.  
The radial distribution comes from the Macdonald function (with $n-1$ nodes
at $r>r_0$),  which explains the sharp drop from a finite value to almost zero
when going outwards from the origin.  Finally, because of the proliferation of
bound states near the gap edges, the total density of states, $\nu(E)$,
becomes singular as $|E|$ approaches $\Delta$ from below, 
\begin{equation}
\nu(E) \simeq \frac{1}{\Delta-|E|} \sum_{j,\kappa} \Theta(p-p_{j,\kappa}) 
\frac{s_{j,\kappa}}{2\pi},
\end{equation} 
with the Heaviside step function $\Theta$.
Every $(j,\kappa)$ tower with $p>p_{j,\kappa}$ here contributes to
the prefactor through the Efimov exponent $s_{j,\kappa}$ in Eq.~(\ref{ssdef}).

\textit{Two-center potential.---}Let us now briefly address the 
two-center potential $V$ in Eq.~(\ref{twocenter}), again for
$p\ll d^2\Delta$ and $\varepsilon\ll \Delta$, where the 
2D Schr\"odinger equation (\ref{schr}) applies.  Using elliptic coordinates
$\xi\ge 1$ and $-1\le\eta\le 1$ \cite{gradst}, where 
$V(\xi,\eta) = 4p\eta/[(\xi^2-\eta^2)d^2]$, 
the problem separates with the ansatz
$\chi(\xi,\eta) =  \frac{Y(\eta)}{(1-\eta^2)^{1/4}} 
\frac{R(\xi)}{(\xi^2-1)^{1/4}}.$
With the separation constant $A=-\gamma+1/4$, the 'angular' and 'radial' 
equations, resp.,  
\begin{eqnarray} \nonumber
&& \left( \frac{d^2}{d\eta^2} + \frac{2p\Delta\eta-A} {1-\eta^2} +
\frac{3/4}{(1-\eta^2)^2} -\frac{\varepsilon\Delta d^2}{2} 
\right) Y(\eta)=0, \\ \label{AKrad}
&& \left( \frac{d^2}{d\xi^2} +\frac{A}{\xi^2-1} +\frac{3/4}{(\xi^2-1)^2} 
-\frac{\varepsilon \Delta d^2}{2} \right) R(\xi)=0,
\end{eqnarray}
coincide with the  Abramov-Komarov equations for 
the 3D Schr\"odinger problem \cite{AK}.  
Adapting their analysis for $p\Delta\gg 1$, 
we find $\gamma<0$ for $p>p_{j,\kappa}^{\rm AK}$ with
\begin{equation}\label{akdef}
p_{j,\kappa}^{\rm AK}\Delta= \frac{\Gamma^4(1/4)}{64\pi} \left[
\left(2j+\frac{\kappa}{2}\right)^2-\frac{1}{6\pi} \right],
\end{equation}
where $j$ and $\kappa$ take the same values as above. 
By construction, Eq.~(\ref{akdef}) is highly accurate for 
$p\Delta\gg 1$, but Table \ref{table11} demonstrates that it
works very well even for $p\Delta\approx 1.9$. 
Not surprisingly, the exact result $p_{0,+}=0$ is not
captured by this approach, $p_{0,+}^{\rm AK}\Delta\simeq 0.17$. However,
$p_{0,+}=0$ follows from an exact calculation 
for the two-center potential \cite{dipol1}. 
Interestingly, Eq.~(\ref{akdef}) also provides an
analytical approximation for the zeroes of the Mathieu characteristic values.
Solving Eq.~(\ref{AKrad}) as in Ref.~\cite{AK},
we recover the spectrum in Eq.~(\ref{spectrum}) with $s_{j,\kappa}$ 
in Eq.~(\ref{ssdef}), where $\alpha=4\pi/\Gamma^2(1/4)$ and $r_0\to d/4$.
While apart from the $(0,+)$ tower, bound state energies are
obtained in accurate analytical form, $|\Psi(x,y)|^2$ is given only 
implicitly and thus difficult to extract.
Finally, the excellent agreement with the point dipole result confirms that 
short-distance regularization issues are irrelevant. 

\begin{figure}
\centering
\includegraphics[width=8.5cm]{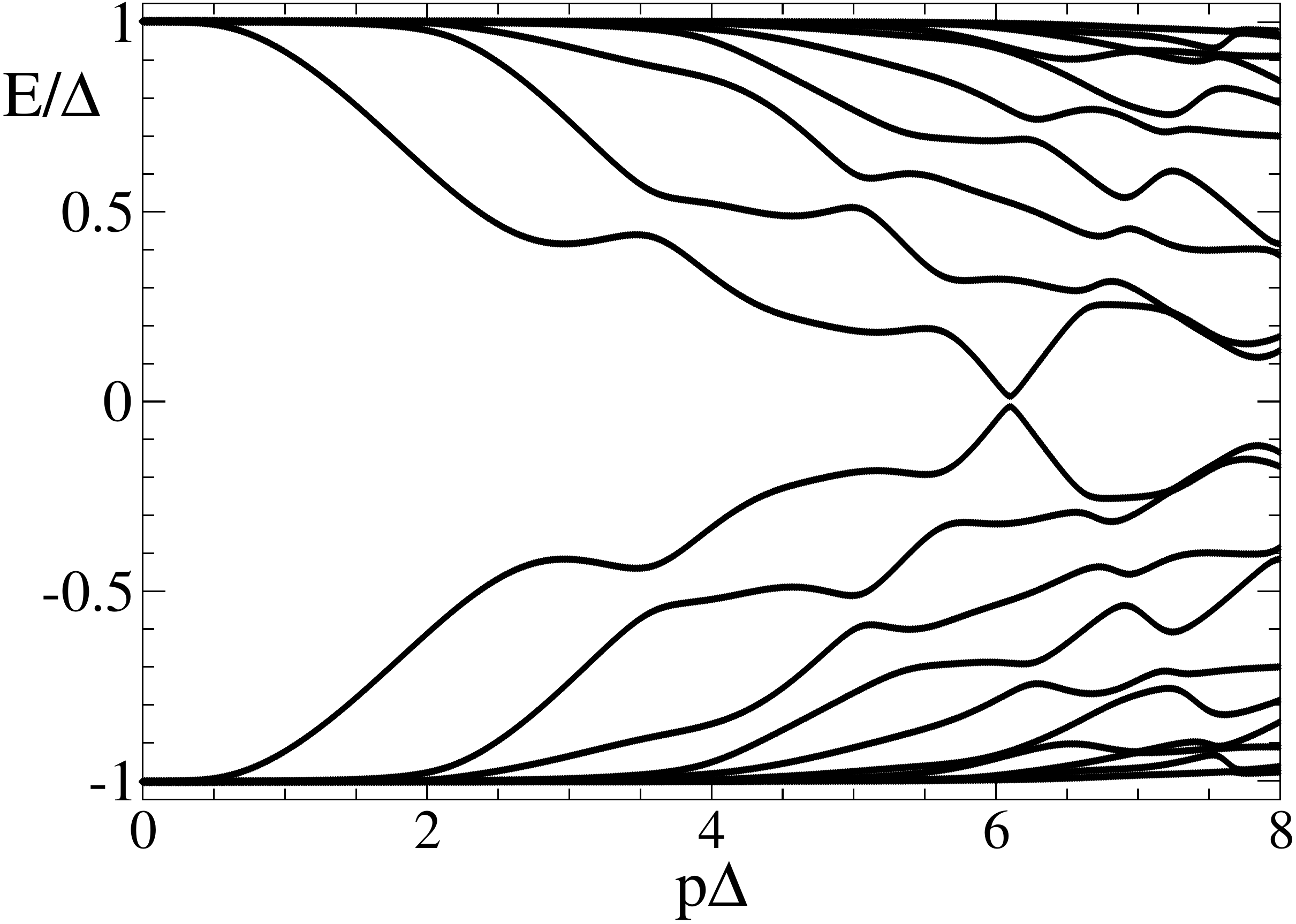}
\caption{\label{fig3} Bound state spectrum vs dipole moment for a circular 
graphene flake with radius $R_{\rm fl}=75r_0$, where  $r_0=d/2$, from
exact diagonalization of Eq.~(\ref{diracdip}) with infinite-mass boundary
conditions at $r=R_{\rm fl}$.  }
\end{figure}

\textit{Numerical diagonalization.---}Since finite-size effects can be 
important in practice, we have studied the bound-state
spectrum for a circular graphene flake of radius $R_{\rm fl}\gg d$, 
using the full Dirac equation (\ref{diracdip})
for the point dipole in Eq.~(\ref{pointdipole}) \cite{foot3}.
We impose infinite-mass boundary conditions \cite{berry} at 
$r=R_{\rm fl}$, which is consistent with the particle-hole
symmetry [Eq.~(\ref{unitary})] and allows us to compute the spectrum
by exact diagonalization, see Fig.~\ref{fig3}. 
For the value of $R_{\rm fl}$ chosen in Fig.~\ref{fig3}, Efimov scaling is not
yet fully developed, but the observed spectrum shows
the emergence of new bound state towers as the dipole moment increases.  
Figure \ref{fig3} also clarifies the fate of bound states upon increasing 
the dipole moment.  First, we find that bound states do not dive into the 
continuum.  This agrees with our analytical results, 
which are exact close to the gap edges, and indicates that supercriticality is 
unlikely to occur. Second, with increasing $p$, 
bound state energies tend to approach (without ever reaching) zero energy.
In fact, the absence of midgap ($E=0$) states can be explained as follows: 
Equation (\ref{unitary}) implies that a putative zero mode must 
be of the form $\Psi_{E=0}(r,\theta)= (\psi(r,\theta), \pm \psi(r,\pi-\theta))^T$,
with a function $\psi(r,\theta)$.  Choosing the $+$ sign (the same follows 
with the $-$ sign) and  $\Delta\to 0$,
the Dirac equation (\ref{diracdip}) reduces to
\begin{equation}
\frac{p\cos\theta}{r^2} \psi(r,\pi-\theta) + e^{i\theta}\left(i\partial_r-
\frac{1}{r}\partial_\theta\right) \psi(r,\theta) =0.
\end{equation}
The radial dependence is solved by  $\psi\sim e^{i(p/r)y(\theta)}$, with 
an angular function $y(\theta)=y(\pi-\theta)$.  However, the resulting equation 
for $y(\theta)$  does not admit a solution.  We conclude that zero modes, given 
their absence for $\Delta\to 0$, are unlikely to exist for finite $\Delta$ \cite{foot4}.

\textit{Scattering states.---}Finally, we 
turn to continuum solutions of the Dirac equation
 with $V$ in Eq.~(\ref{twocenter}). For simplicity, we consider $|E|\gg \Delta$,
where the Born approximation \cite{novikov,zazu} is applicable.
For an incoming plane wave with momentum ${\bf k}$ and 
$\sigma={\rm sgn}(E)=\pm$, the asymptotic scattering state is \cite{novikov}
\begin{eqnarray} \label{scatt}
 \Psi_{{\bf k},\sigma}(r,\theta) &\simeq &
  e^{ i{\bf k}\cdot {\bf r}} U_{{\bf k},\sigma} 
+ f(\theta,\phi_{\bf k}) \frac{e^{ikr}}{\sqrt{-ir}} U_{{\bf k}',\sigma} ,
\end{eqnarray}
with ${\bf k}'=k \hat {\bf r}$,  $\phi_{\bf k}$ the angle
between ${\bf k}$ and the dipole ($x$-)axis, and  
$U_{{\bf k},\sigma}  = \frac{1}{\sqrt{2}} 
\left(\begin{array}{c} e^{-i\phi_{\bf k}/2}\\ \sigma  
e^{i\phi_{\bf k}/2} \end{array} \right)$.
For long wavelengths, $kd\ll 1$, the scattering amplitude is
\begin{equation}\label{scattamp}
f(\theta,\phi_{\bf k}) \simeq i p\sqrt{2\pi k} 
\cos[(\theta-\phi_{\bf k})/2] \sin[(\theta+\phi_{\bf k})/2]. 
\end{equation}
The transport and total cross-sections, 
$\Lambda_{\rm tr}=\int d\theta [1-\cos(\theta-\phi_{\bf k})] |f(\theta)|^2$ 
and $\Lambda =\int d\theta |f(\theta)|^2$ \cite{novikov},
resp., are then given by 
$\Lambda_{\rm tr}= \frac{\pi^2}{2} p^2 k$
and
$ \Lambda = \left(1+2\sin^2\phi_{\bf k}\right)  \Lambda_{\rm tr}$.
Remarkably, $\Lambda_{\rm tr}$ is independent of $\phi_{\bf k}$,
with the dipole-induced angular dependence precisely compensated by the 
$\cos[(\theta-\phi_{\bf k})/2]$ factor in Eq.~(\ref{scattamp}). This factor
is specific for Dirac fermions and causes the well-known
'absence of backscattering' by short-ranged impurities \cite{rmp1}.
We then expect the electrical conductivity of a graphene sample containing 
oriented dipoles to be \textit{isotropic}. 

\textit{Conclusions.---}The electric dipole problem for 2D Dirac fermions 
exhibits rich physics that could be probed by STM spectroscopy in graphene.  
The Efimov-like scaling of the bound state energies, with 
the gap edges as accumulation points, suggests that electrons can be 
captured (and thus confined) by a dipole potential.  This scaling
property, formally identical to the scaling of the three-body
levels of identical bosons, here emerges in a different physical setting
and can be traced to the $1/r^2$ dependence of the dipole potential.
While we have disregarded electron-electron interactions 
beyond a Fermi velocity renormalization  \cite{rmp2},
$\Delta$ tends to suppress charge fluctuations and no
profound changes are expected for weak interactions.  
Future work should clarify whether multi-electron bound states are
possible in such a setting.  
  
\textit{Acknowledgments.}---We thank A. Altland, E. Andrei, H. Siedentop, 
and A. Zazunov for discussions, and the DFG (SFB TR12 and SPP 1459) 
and the Volkswagen-Stiftung  for financial support.


\begin{thebibliography}{99}
\bibitem{rmp1} A.H. Castro Neto, F. Guinea, N.M.R. Peres, K.S. Novoselov, and A. Geim, Rev. Mod. Phys. {\bf 81}, 109 (2009).
\bibitem{strain} M.A.H. Vozmediano, M.I. Katsnelson, and F. Guinea, Phys. Rep. {\bf 496}, 109 (2010).
\bibitem{soi} D. Huertas-Hernando, F. Guinea, and A. Brataas, Phys. Rev. B {\bf 74}, 155426 (2006).
\bibitem{rmp2} V.N. Kotov, B. Uchoa, V.M. Pereira, A.H. Castro Neto, and F. Guinea, Rev. Mod. Phys. {\bf 84}, 1067 (2012).
\bibitem{pono} L.A. Ponomarenko, R.V. Gorbachev, G.L. Yu, D.C. Elias, R. Jalil, A.A. Patel, A. Mishchenko, A.S. Mayorov, C.R. Woods, J.R. Wallbank, M. Mucha-Kruczynski, B.A. Piot,
M. Potemski, I.V. Grigorieva, K.S. Novoselov, F. Guinea, V.I. Fal'ko, and A.K. Geim,
Nature (London) {\bf 497}, 594 (2013).
\bibitem{levitov2} J.C.W. Song, A.V. Shytov, and L.S. Levitov, Phys. Rev.  Lett. {\bf 111}, 266801 (2013). 
\bibitem{novikov} D.S. Novikov, Phys. Rev. B {\bf 76}, 245435 (2007).
\bibitem{pereira} V.M. Pereira, J. Nilsson, and A.H. Castro Neto, Phys. Rev.  Lett. {\bf 99}, 166802 (2007).
\bibitem{levitov} A.V. Shytov, M.I. Katsnelson, and L.S. Levitov, Phys. Rev. Lett. {\bf 99}, 246802 (2007).
\bibitem{gamayun} O.V. Gamayun, E.V. Gorbar, and V.P. Gusynin, Phys. Rev. B {\bf 80}, 165429 (2009).
\bibitem{denis} D. Kl\"opfer, A. De Martino, and R. Egger, Crystals {\bf 3}, 14 (2013).
\bibitem{exp1} Y. Wang, V.W. Brar, A.V. Shytov, Q. Wu, W. Regan, H.-Z. Tsai, A. Zettl, L.S. Levitov, and M.F. Crommie, Nat. Phys.  {\bf 8}, 653 (2012). 
 \bibitem{exp2} A. Luican-Mayer, M. Kharitonov, G. Li, C.P. Lu, 
I. Skachko, A.M.B. Goncalves, K. Watanabe, T. Taniguchi, and E.Y. Andrei, 
Phys. Rev. Lett. {\bf 112}, 036804 (2014).
\bibitem{exp3} Y. Wang, D. Wong, A.V. Shytov, V.W.  Brar, S. Choi, Q. Wu, H.-Z. Tsai, W. Regan, A. Zettl, R.K. Kawakami, S.G. Louie, L.S. Levitov, and M.F. Crommie, Science {\bf 340}, 734 (2013).  
\bibitem{wrinkle} V.M. Pereira, A.H. Castro Neto, H.Y. Liang, and L. Mahadevan, Phys. Rev. Lett. {\bf 105}, 156603 (2010).
\bibitem{greiner} W. Greiner, B. M\"uller, and J. Rafelski, \textit{Quantum electrodynamics of Strong Fields} (Springer, Berlin, 1985).
\bibitem{popov} V.S. Popov, Phys. At. Nucl. {\bf 64}, 367 (2001).
\bibitem{efimov} V. Efimov, Phys. Lett. B {\bf 33}, 563 (1970).
\bibitem{efimov2} E. Braaten and H.W. Hammer, Phys. Rep. {\bf 428}, 259 (2007).
\bibitem{gogolin} A.O. Gogolin, C. Mora, and R. Egger, 
Phys. Rev.  Lett. {\bf 100}, 140404 (2008).
\bibitem{gomes} K.K. Gomes, W. Mar, W. Ko, F. Guinea, and H.C. Manoharan, Nature (London) {\bf 483}, 306 (2012). 
\bibitem{hasan} M.Z. Hasan and C.L. Kane, Rev. Mod. Phys. {\bf 82}, 3045 (2010).
\bibitem{foot1}
A substrate dielectric constant can be included by renormalization of $p$. 
The potential (\ref{pointdipole}) also arises by deposition of a 
polar molecule. 
\bibitem{gusynin} O.O. Sobol, E.V. Gorbar, and V.P. Gusynin, Phys. Rev. B  {\bf 88}, 205116 (2013).
\bibitem{popov2} A.A. Perelomov and V.S. Popov, Theor. Math. Phys. {\bf 4}, 664 (1970).
\bibitem{AK} D.I. Abramov and I.V. Komarov, Theor. Math. Phys.  {\bf 13}, 209 (1972).  
\bibitem{davron} D.U. Matrasulov, V.I. Matveev, and M.M. Musakhanov, Phys. Rev. A {\bf 60}, 4140 (1999).
\bibitem{dipol0} H.E. Camblong, L.N. Epele, H. Fanchiotti, and C.A.G. Canal, Phys. Rev. Lett. {\bf 87}, 220402 (2001).
\bibitem{dipol1} K. Connolly and D.J. Griffiths, Am. J. Phys. {\bf 75}, 527 (2007).
\bibitem{num-efimov} D. Schumayer, B.P. Zyl, R.K. Bhadure, and D.A.W. Hutchinson, EPL {\bf 89}, 13001 (2010).
\bibitem{gradst} I.S. Gradshteyn and I.M. Ryzhik, \textit{Table of Integrals, Series, and Products} (Academic Press, Elsevier, 2007).
\bibitem{abramowitz} M. Abramowitz and I.A. Stegun (eds.), \textit{Handbook of Mathematical Functions} (Dover, New York, 1965).  
\bibitem{foot2}
On the level of the Dirac equation, 
this corresponds to vanishing radial current at $r=r_0$.
\bibitem{foot3} 
In the numerics, we use the regularization $V(r<r_0)=-p\cos(\theta)/r_0^2$.
\bibitem{berry} M.V. Berry and R.J. Mondragon, Proc. R. Soc. London A {\bf 412}, 53 (1987).
\bibitem{foot4}
For finite $\Delta$, the squared Dirac equation for $E=0$ does not contain
terms $\propto 1/r^2$ for $r\to \infty$, but at most terms $\propto 1/r^3$
which do not allow for bound states.  
\bibitem{zazu} A. Zazunov, A. Kundu, A. H\"utten, and R. Egger, Phys. Rev. B {\bf 82}, 155431 (2010).
\end{thebibliography}
\end{document}